\documentclass[superscriptaddress,twocolumn,showpacs,a4paper,
amssymb,amsmath,nobibnotes,aps,prd,
showkeys,
nofootinbib,notitlepage]{revtex4-1}
\usepackage{verbatim}
\usepackage[T1]{fontenc}
\usepackage[utf8]{inputenc}
\usepackage[american]{babel}
\usepackage{epsfig}
\usepackage{graphicx,subcaption,caption}
\usepackage{booktabs}
\usepackage{multirow}
\usepackage{dcolumn}
\usepackage{amsmath}
\usepackage{mathtools}
\usepackage{amsfonts}
\usepackage{amssymb}
\usepackage{ulem}
\usepackage{epstopdf}
\usepackage{bm}
\usepackage{siunitx}
\usepackage{braket}
\usepackage{enumitem}
\usepackage{soul}
\usepackage[table]{xcolor}
\usepackage{color}
\usepackage{transparent}
\usepackage{pifont}
\usepackage{enumitem}

\definecolor{navyblue}{rgb}{0.0, 0.0, 0.5}
\definecolor{royalblue}{rgb}{0.25, 0.41, 0.88}
\definecolor{cadmiumgreen}{rgb}{0.0, 0.42, 0.24}
\definecolor{blue-violet}{rgb}{0.54, 0.17, 0.89}
\definecolor{darkviolet}{rgb}{0.58, 0.0, 0.83}
\definecolor{orange(colorwheel)}{rgb}{1.0, 0.5, 0.0}

\usepackage{hyperref}
\hypersetup{
    colorlinks=true, 
    linkcolor=royalblue, 
    citecolor=magenta}

\usepackage{booktabs}
\usepackage{multirow}
\usepackage{dcolumn}
\usepackage{colortbl}

\begin{document}

\title{Lensing impact on cosmic relics and tensions}
\date{\today}

\author{William Giar\`e}
\email{w.giare@sheffield.ac.uk}
\affiliation{Consortium for Fundamental Physics, School of Mathematics and Statistics, University of Sheffield, Hounsfield Road, Sheffield S3 7RH, United Kingdom}
\author{Olga Mena}
\email{omena@ific.uv.es}
\affiliation{Instituto de F{\'\i}sica Corpuscular  (CSIC-Universitat de Val{\`e}ncia), E-46980 Paterna, Spain}
\author{Eleonora Di Valentino}
\email{e.divalentino@sheffield.ac.uk}
\affiliation{School of Mathematics and Statistics, University of Sheffield, Hounsfield Road, Sheffield S3 7RH, United Kingdom}

\begin{abstract}
Cosmological bounds on neutrinos and additional hypothetical light thermal relics, such as QCD axions, are currently among 
the most restrictive ones. These limits mainly rely on Cosmic Microwave Background temperature anisotropies. Nonetheless, one of the largest cosmological signatures of thermal relics is that on gravitational lensing, due to their free streaming behavior before their non-relativistic period. We investigate \textit{late time only} hot relic mass constraints, primarily based on recently released lensing data from the Atacama Cosmology Telescope, both alone and in combination with lensing data from the Planck Satellite. Additionally, we consider other local probes, such as Baryon Acoustic Oscillations measurements, shear-shear, galaxy-galaxy, and galaxy-shear correlation functions from the Dark Energy Survey, and distance moduli measurements from Type Ia Supernovae. The tightest bounds we find are $\sum m_\nu<0.43$~eV and $m_a<1.1$~eV, both at $95\%$~CL. 
Interestingly, these limits are still much stronger than those found on e.g. laboratory neutrino mass searches, reassessing the robustness of the extraction of thermal relic properties via cosmological observations. In addition, when considering lensing-only data, the significance of the Hubble constant tension is considerably reduced, while the clustering parameter $\sigma_8$ controversy is completely absent.   
\end{abstract}

\maketitle

\section{Introduction}
\label{sec.intro}

In the standard $\Lambda$CDM cosmological model, the three flavors of light active neutrinos, as predicted within the framework of the Standard Model (SM) of elementary particles, contribute to the overall energy density of the Universe as hot thermal relics. However, theoretical attempts to address the deficiencies of the SM of particle physics often introduce additional light and elusive degrees of freedom, leading to a range of new particle candidates for physics beyond the standard model that often exhibit behaviors similar to neutrinos and can also contribute to the overall energy density as thermal relics. 

Among these candidates, the QCD axion~\cite{Peccei:1977hh,Peccei:1977ur,Wilczek:1977pj,Weinberg:1977ma} has gained significant attention in the quest for physics beyond the SM due to its potential implications for the cosmological energy budget and its role in addressing fundamental puzzles in particle physics~\cite{Preskill:1982cy,Abbott:1982af,Dine:1982ah,Baker:2006ts,Pendlebury:2015lrz,Abel:2020pzs}. Axions can be abundantly produced in the early Universe through a wide range of physical mechanisms~\cite{Kibble:1976sj,Vilenkin:1981kz,Kibble:1982dd,Sikivie:1982qv,Vilenkin:1982ks,Linde:1985yf,Huang:1985tt,Seckel:1985tj,Davis:1986xc,Lyth:1989pb,Linde:1990yj,Vilenkin:2000jqa,Turner:1986tb,Berezhiani:1992rk,Brust:2013ova,Baumann:2016wac,DEramo:2018vss,Arias-Aragon:2020qtn,Arias-Aragon:2020shv,Green:2021hjh,DEramo:2021usm} and the implications of a cosmic axion background depend on the specific production mechanism employed, see \textit{e.g.}, Ref.~\cite{DiLuzio:2020wdo} for a recent review. If axions are thermally produced through scatterings and particle decays within the primordial bath, they contribute to the radiation energy-density, similar to massive neutrinos, and can be classified as a component of hot dark matter~\cite{Turner:1986tb,Berezhiani:1992rk,Brust:2013ova,Baumann:2016wac,DEramo:2018vss,Arias-Aragon:2020qtn,Arias-Aragon:2020shv,Green:2021hjh,DEramo:2021usm,Hannestad:2005df,Melchiorri:2007cd,Hannestad:2007dd,Hannestad:2008js,Hannestad:2010yi,Archidiacono:2013cha,Giusarma:2014zza,DiValentino:2015zta,DiValentino:2015wba,Archidiacono:2015mda,DEramo:2021psx,DEramo:2021lgb}.

Cosmology can set very strong bounds on hot dark matter relics, including standard neutrinos~\cite{DiValentino:2021hoh,Palanque-Delabrouille:2019iyz,diValentino:2022njd} and the hypothesized thermal axions~\cite{Giare:2020vzo,Giare:2021cqr,DEramo:2022nvb,Notari:2022ffe,DiValentino:2022edq}. As for neutrinos, cosmological observations currently provide the tightest limit on the total neutrino mass, $\sum m_\nu<0.09$~eV at $95\%$~CL~\cite{DiValentino:2021hoh,Palanque-Delabrouille:2019iyz,diValentino:2022njd}. This constraint is comparable to the Inverted Ordering  lower bound ($\sum m_\nu>0.0997 \pm 0.00051$~eV)  derived from neutrino oscillation data~\cite{deSalas:2020pgw,Esteban:2020cvm,Capozzi:2021fjo}. Concerning axions, the most constraining bound in the literature in a mixed hot dark matter cosmology is $m_a\lesssim 0.2$~eV, together with limits on the neutrino sector of  $\Delta N_{\rm eff} < 0.23$ and $\sum m_\nu < 0.16$~ eV, all at $95\%$~CL~\cite{Giare:2020vzo,Giare:2021cqr,DEramo:2022nvb,Notari:2022ffe,DiValentino:2022edq}.

All the limits above mainly rely on early Universe observations such as the Planck Cosmic Microwave Background (CMB) data or Big Bang Nucleosynthesis (BBN) measurements. Indeed it is well known that thermal relics play a non-negligible role both during the BBN epoch and in the CMB temperature anisotropies. For instance, during the decoupling epoch, light neutrinos can transit from a relativistic to a non-relativistic regime, thereby impacting the gravitational potentials and leaving characteristic signatures in the angular power spectra of temperature and polarization anisotropies through the integrated Sachs-Wolfe (ISW) effect, depending this imprint  on the total neutrino mass $\sum m_{\nu}$. However, this effect, as well as the horizontal shift towards larger angular scales induced in the CMB temperature anisotropies, is largely degenerate with other cosmological parameters, as, for instance, the Hubble constant $H_0$~\cite{Bond:1997wr,Zaldarriaga:1997ch,Efstathiou:1998xx}.  When substituting Planck data with a combined analysis of other independent CMB measurements provided by WMAP, ACT, and SPT-3G, the previously mentioned constraints are generally relaxed~\cite{ACT:2020gnv,SPT-3G:2021eoc,DiValentino:2022oon}. In certain cases, the neutrino mass bounds exceed the eV limit\footnote{Notice however that it can be brought down to $\sum m_{\nu}\lesssim 0.2$ eV when including large scale structure information~\cite{DiValentino:2023fei}.} and even show a slight preference for a larger neutrino mass value~\cite{DiValentino:2021imh}. This underscores the importance of obtaining complementary constraints on thermal relics that are independent of CMB anisotropies, as they can serve as a valuable source for cross-validation.\footnote{This discrepancy present in the total neutrino mass constraints has been investigated for several extensions of the $\Lambda$CDM model, revealing a CMB tension between Planck and ACT~\cite{Handley:2020hdp,Giare:2022rvg,DiValentino:2022rdg,DiValentino:2022oon,Forconi:2021que,Giare:2023wzl,Hill:2021yec,Poulin:2021bjr}.}

In this regard, Planck observations opened up a new era in which the  the dominant effect of neutrinos is due to \textit{gravitational lensing}~\cite{Planck:2013pxb}. 
After decoupling from the thermal bath, neutrinos travel without interactions along geodesics as hot thermal relics with significant velocity dispersions. The non-relativistic neutrino overdensities only cluster at wavelengths larger than their free streaming scale, hindering the growth of matter fluctuations on small scales and suppressing galaxy clustering and leaving a distinct imprint on the lensing potential, especially on scales smaller than the horizon when they become non-relativistic.  Increasing neutrino masses will increase the expansion rate at $z \gtrsim 1$, suppressing clustering on scales smaller than the horizon size at the non relativistic transition~\cite{Kaplinghat:2003bh,Lesgourgues:2005yv}. This translates into a supression of the CMB lensing power spectrum of $10\%$ at multipoles $\ell=1000$, assuming the minimum neutrino mass indicated by neutrino oscillation experiments, i.e. $\sum m_\nu <0.06$~eV~\cite{Planck:2013pxb}.

 The former significant impact of thermal relics in the late Universe, corroborated by the significant improvement in the constraints on their properties that are achieved exploiting observations of the local Universe, suggests that ongoing advancements in reconstructing the dark matter distribution, particularly through the lensing spectrum, combined with other precise cosmological observations in the late-time Universe, may offer a promising approach to constrain thermal relics based solely on their indirect effects at later times.  In this regard, the recent data Release 6 (DR6) from the Atacama Cosmology Telescope~\cite{ACT:2023kun,ACT:2023dou} has provided a comprehensive reconstruction of Cosmic Microwave Background lensing over 9400 sq. deg. of the sky, opening up new avenues for studying the properties of neutrinos and other light particles.\footnote{Remarkably, when combined with Planck CMB anisotropies data and Baryon Acoustic Oscillation measurements, this dataset yields an upper bound of $\sum m_{\nu} < 0.12$ eV. This upper bound remains unchanged even when incorporating the Planck satellite lensing data.}
 
 In this work, we aim to examine the constraints that can be obtained on thermal relics solely from observations of the local universe, particularly focusing on the impact derived from recent lensing measurements in combination with other large-scale structure data. The paper is structured as follows: In \autoref{sec.method}, we will describe the methodology used in our data analysis. In \autoref{sec.results}, we present the main results, distinguishing between three possible scenarios. We begin by studying the simplest and most typical case where all thermal relics are represented by massive neutrinos only (see~\autoref{sec.neutrinos}). Next, we fix the neutrino mass to the reference value of $\sum m_{\nu} \sim 0.06$ eV and analyze the constraints that can be achieved on thermal axions (\autoref{sec.axion}). Finally, we perform a full joint analysis of axions and neutrinos (see \autoref{sec.joint}). In \autoref{sec.conclusion} we draw our conclusions.

\section{Numerical Methodology}
\label{sec.method}

\subsection{Thermal Relics Implementation}

We investigate the effects of relic populations of neutrinos and thermal axions at late cosmic times by employing a modified version of the Boltzmann integrator code \texttt{CAMB}~\cite{Lewis:1999bs,Howlett:2012mh}. Our tailored modifications comprehensively consider the effects of QCD axions across various cosmological scales and epochs by incorporating the axion mass as an additional cosmological parameter, similarly to massive neutrinos.\footnote{It is important to note that our code has been extensively tested in previous studies and is built upon a strong understanding of thermal axions across the QCD phase transition~\cite{DEramo:2021psx,DEramo:2021lgb}. In addition, our previous results have been verified to match those obtained by independent groups using different numerical methodologies (see for instance Refs.~\cite{DiLuzio:2021vjd,DiLuzio:2022gsc,DEramo:2022nvb,DiValentino:2022edq,Notari:2022ffe,Caloni:2022uya}). We refer to Ref.~\cite{DEramo:2022nvb} for a detailed step-by-step explanation of all the modifications involved.}

Specifically, our code allows us to disentangle the distinct effects of QCD axions during early and late cosmic times. In the early Universe, when the axion is relativistic, it behaves as radiation and contributes to the effective number of neutrino species $N_{\rm eff}$. To accurately compute this contribution, we solve the Boltzmann equation for the axion number density, specifically focusing on the KSVZ model of axion-hadron interactions~\cite{Kim:1979if,Shifman:1979if}. While the axion is relativistic, it directly impacts the CMB angular power spectra through the early integrated Sachs-Wolfe effect, similar to massive neutrinos. Additionally, it indirectly modifies the primordial helium abundance during Big Bang Nucleosynthesis (BBN). However, since our analysis does not incorporate observations of the early Universe, these effects have minimal impact on our results, except for their potential indirect influence on the lensing spectrum.

On the other hand, as the axion switches to a non-relativistic regime, it behaves as cold dark matter, leading a significant influence on the process of structure formation. Notably, depending on the value of its mass, the axion may become non-relativistic much earlier than massive neutrinos. This characteristic enables us to distinguish between the effects left by massive neutrinos and massive axions on structure formation,\footnote{Nonetheless, when the masses of axions and neutrinos are similar, the evolution of their energy densities hinders our ability to constrain their masses to values below $\sim 0.1$ eV, as discussed in Refs.~\cite{Giare:2021cqr,DEramo:2022nvb}.} thereby offering us the opportunity to derive joint constraints on thermal relics solely through the analysis of the lensing potential reconstruction and large-scale structure data.

\subsection{Markov Chain Monte Carlo Analysis}

In order to derive observational constraints on hot thermal relics from late-time data, we perform Markov Chain Monte Carlo (MCMC) analyses using the publicly available version of the sampler \texttt{COBAYA}~\cite{Torrado:2020xyz}. As our primary focus is on lensing measurements in combination with other late times observations, we use the identical setup and assumptions employed by the ACT collaboration in their lensing Data Release 6 paper on cosmological parameters, Ref.~\cite{ACT:2023kun}. Specifically, we adopt the same likelihoods set-up discussed in Appendix A of Ref.~\cite{ACT:2023kun}, considering a fiducial cosmological model that extends the $\Lambda$CDM model to include hot relics such as neutrinos and/or axions. To ensure a direct comparison of our results with those obtained by the ACT collaboration, we also adopt the same priors on the standard cosmological parameters, as presented in Table 1 of Ref.~\cite{ACT:2023kun}, where the CMB anisotropies data are excluded. We summarize these priors in~\autoref{tab.Priors}.

The convergence of the chains obtained with this procedure is tested using the Gelman-Rubin criterion~\cite{Gelman:1992zz}, and we choose as a threshold for chain convergence $R-1 \lesssim 0.02$.

\begin{table}[t!]
	\begin{center}
		\renewcommand{\arraystretch}{1.5}
		\begin{tabular}{l@{\hspace{2 cm}} c}
			\hline
			\textbf{Parameter}    & \textbf{Prior} \\
			\hline\hline
			$\Omega_{\rm c} h^2$         & $[0.005\,,\,0.99]$ \\
			$\Omega_{\rm b} h^2$         & $\mathcal N(0.02233\,,\,0.00036)$ \\
			$100\,\theta_{\rm {MC}}$     & $[0.5\,,\,10]$ \\
			$\log(10^{10}A_{\rm S})$     & $[1.61\,,\,4]$ \\
			$n_{\rm s}$                  & $\mathcal N(0.96,0.02)$ \\
			$\sum m_{\nu}$ [eV]          & $[0.06\,,\,5]$\\
			$m_{\rm a}$ [eV]             & $[0.01\,,\,10]$\\
			\hline\hline
		\end{tabular}
          \caption{\small Prior distributions adopted for cosmological parameters. Uniform priors are shown in square brackets and Gaussian priors with mean $\mu$ and standard deviation $\sigma$ are shown as $\mathcal N (\mu\,,\,\sigma)$.}
	\label{tab.Priors}
	\end{center}
\end{table}

\subsection{Datasets}
The reference datasets exploited in our analysis are:

\begin{itemize}

\item  The gravitational lensing mass map covering 9400 deg$^2$
reconstructed from measurements of the Cosmic Microwave Background made by the Atacama Cosmology Telescope from 2017 to 2021 \cite{ACT:2023kun,ACT:2023dou}. In our analysis we include only the conservative range of lensing multipoles $40 < \ell < 763$. This dataset is referred to as \textbf{\textit{ACT-DR6}}.

\item The ACT-DR6 dataset is considered both independently and in conjunction with the \textit{Planck-2018} lensing likelihood~\cite{Planck:2018lbu}, derived from the temperature 4-point correlation function. We shall refer to the combined likelihood which includes data from both ACT-DR6 and \textit{Planck-2018} lensing simply as \textbf{\textit{Full lensing}}.

\end{itemize}

Additionally, we investigate a wide array of local probes of the Universe by combining these two datasets with other late-time observations, including:

\begin{itemize}

\item Baryon Acoustic Oscillations (BAO) and Redshift Space Distortions (RSD) measurements obtained from a combination of the spectroscopic galaxy and quasar catalogues of the Sloan Digital Sky Survey (SDSS)~\cite{BOSS:2012dmf} and the more recent eBOSS DR16 data~\cite{BOSS:2016wmc,eBOSS:2020yzd}. To remain conservative, we exclude Quasars and Lyman-$\alpha$ BAO measurements from this dataset. This likelihood is referred to as \textit{\textbf{BAO}}.

\item The Pantheon catalogue which includes a collection of 1048 B-band observations of the relative magnitudes of Type Ia supernovae~\cite{Pan-STARRS1:2017jku}. This dataset is referred to as \textit{\textbf{SN}}.

\item The shear-shear, galaxy-galaxy, and galaxy-shear correlation functions from the first year of the Dark Energy Survey~\citep{DES:2017myr}. We refer to this dataset as \textit{\textbf{DES}}.

\end{itemize}

\section{Results}
\label{sec.results}


\begin{table*}
\begin{center}
\renewcommand{\arraystretch}{1.5}
\resizebox{\textwidth}{!}{
\begin{tabular}{l c c c c c c c c c c c c c c c }
\hline
\textbf{Parameter} & \textbf{ ACT-DR6 } & \textbf{ ACT-DR6 + BAO } & \textbf{ ACT-DR6 + BAO + DES } & \textbf{ ACT-DR6 + BAO + SN } & \textbf{ ACT-DR6 + BAO + DES + SN } \\ 
\hline\hline

$ \sum m_\nu \, [eV]  $ & $ < 3.32 $ & $ < 1.10 $ & $ < 0.773 $ & $ < 0.717 $ & $ < 0.722 $ \\ 
$ \Omega_\mathrm{m}  $ & $  1.27^{+0.79}_{-1.6}\, (< 3.90 ) $ & $  0.344^{+0.031}_{-0.035}\, ( 0.344^{+0.067}_{-0.063} ) $ & $  0.303\pm 0.014\, ( 0.303^{+0.030}_{-0.029} ) $ & $  0.316\pm 0.016\, ( 0.316^{+0.035}_{-0.034} ) $ & $  0.302\pm 0.012\, ( 0.302^{+0.023}_{-0.021} ) $ \\  
$ H_0  $ & $ < 58.8 $ (unc) & $ 68.7\pm 1.4\, ( 68.7^{+3.0}_{-2.9} ) $ & $  67.21\pm 0.87\, ( 67.2^{+1.8}_{-1.7} ) $ & $  67.9\pm 1.1\, ( 67.9^{+2.2}_{-2.0} ) $ & $  67.16\pm 0.87\, ( 67.2^{+1.7}_{-1.7} ) $ \\ 
$ \sigma_8  $ & $  0.62\pm 0.16\, ( 0.62^{+0.29}_{-0.29} ) $ & $  0.796\pm 0.019\, ( 0.796^{+0.038}_{-0.038} ) $ & $  0.778\pm 0.018\, ( 0.778^{+0.034}_{-0.034} ) $ & $  0.797\pm 0.020\, ( 0.797^{+0.037}_{-0.040} ) $ & $  0.779\pm 0.017\, ( 0.779^{+0.032}_{-0.032} ) $ \\ 

\hline \hline
\end{tabular} }
\end{center}
\caption{\small \textbf{Neutrinos:} Mean values for some of the most relevant cosmological parameters in this study, together with their 68$\%$ (95$\%$) CL errors for the some of the possible data combinations here considered, based on the baseline ACT-DR6 lensing dataset. Upper bounds are quoted at $95\%$~CL significance.}
\label{tab:nu}
\end{table*}

\begin{table*}
\begin{center}
\renewcommand{\arraystretch}{1.5}
\resizebox{\textwidth}{!}{
\begin{tabular}{l c c c c c c c c c c c c c c c }
\hline
\textbf{Parameter} & \textbf{ Full lensing } & \textbf{ Full lensing + BAO } & \textbf{ Full lensing + BAO + DES } & \textbf{ Full lensing + BAO + SN } & \textbf{ Full lensing + BAO + DES + SN } \\ 
\hline\hline

$ \sum m_\nu \, [eV]  $ & $ < 1.42 $ & $ < 0.527 $ & $ < 0.664 $ & $ < 0.490 $ & $ < 0.606 $ \\ 
$ \Omega_\mathrm{m}  $ & $  0.55^{+0.13}_{-0.15}\, ( 0.55^{+0.29}_{-0.27} ) $ & $  0.320\pm 0.010\, ( 0.320^{+0.023}_{-0.022} ) $ & $  0.316\pm 0.011\, ( 0.316^{+0.024}_{-0.023} ) $ & $  0.3173\pm 0.0094\, ( 0.317^{+0.021}_{-0.020} ) $ & $  0.3130\pm 0.0097\, ( 0.313^{+0.021}_{-0.020} ) $ \\ 
$ H_0  $ & $  55.2^{+5.5}_{-6.2}\, ( 55^{+10}_{-10} ) $ & $  67.90\pm 0.73\, ( 67.9^{+1.4}_{-1.4} ) $ & $  68.29\pm 0.71\, ( 68.3^{+1.4}_{-1.4} ) $ & $  68.07\pm 0.72\, ( 68.1^{+1.4}_{-1.5} ) $ & $  68.38\pm 0.71\, ( 68.4^{+1.4}_{-1.4} ) $ \\ 
$ \sigma_8  $ & $  0.689\pm 0.053\, ( 0.69^{+0.11}_{-0.11} ) $ & $  0.796\pm 0.017\, ( 0.796^{+0.035}_{-0.036} ) $ & $  0.776\pm 0.017\, ( 0.776^{+0.036}_{-0.037} ) $ & $  0.798\pm 0.016\, ( 0.798^{+0.033}_{-0.034} ) $ & $  0.779\pm 0.016\, ( 0.779^{+0.033}_{-0.034} ) $ \\

\hline \hline
\end{tabular} }
\end{center}
\caption{ \small  \textbf{Neutrinos:} Mean values for some of the most relevant cosmological parameters in this study, together with their 68$\%$ (95$\%$) CL errors for the some of the possible data combinations here considered, based on the baseline ACT-DR6 plus Planck lensing datasets. Upper bounds are quoted at $95\%$~CL significance.}
\label{tab:nututto}
\end{table*}

\begin{figure*}[t]
    \begin{tabular}{cc}
   \includegraphics[width=0.5\textwidth]{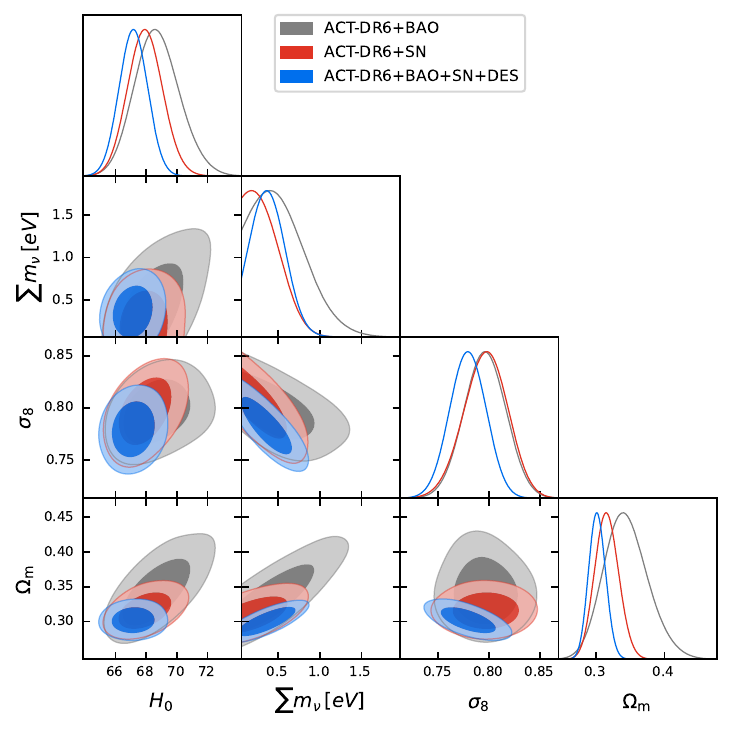}&
   \includegraphics[width=0.5\textwidth]{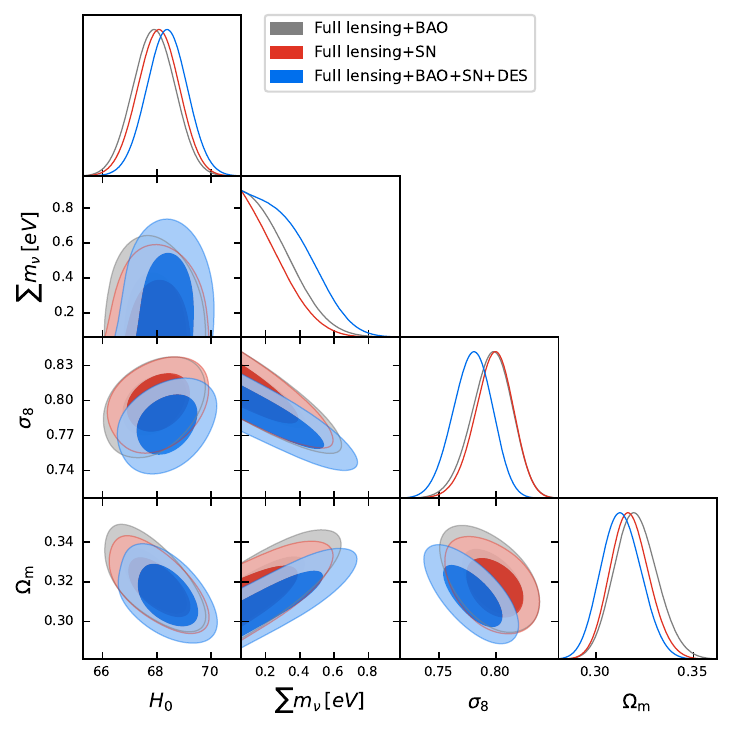}
    \end{tabular}
   \caption{\small \textbf{Neutrinos:} Left (right) panel: One-dimensional posterior probability distributions and two-dimensional $68\%$ and $95\%$~CL allowed contours for $\sum m_a$, $H_0$, $\Omega_{\rm m}$ and $\sigma_8$ from combinations of the baseline ACT-DR6 (ACT-DR6 plus Planck lensing) with other low redshift observables considered along this work.}
    \label{fig:nu}
\end{figure*}

\subsection{Constraints on Neutrinos}
\label{sec.neutrinos}

The main results are summarized in \autoref{tab:nu} and \autoref{tab:nututto} and depicted in~\autoref{fig:nu}. 

\autoref{tab:nu} contains as the basic dataset the ACT-DR6 lensing one. 
Notice that the cosmological constraints by this data set alone are very poor, but nevertheless we show them for completeness. When BAO measurements are included, a $95\%$~CL upper bound on the total neutrino mass of $1.10$~eV is achieved. This result is very remarkable, as it does not rely on a large number of datasets and relies only on large scale and CMB lensing observables. When adding further lensing observations from galaxy surveys such as the one considered along this manuscript (DES), the limit is further strengthened to $0.77$~eV. Nevertheless DES measurements are not very effective when constraining neutrino masses because they prefer a lower value of the clustering parameter $\sigma_8$, which is anti-correlated with the neutrino mass as can be clearly noticed from the contours illustrated in~\autoref{fig:nu}. Indeed, notice that, while the value of $\sigma_8$ from the combination of ACT-DR6 and BAO remains unchanged when adding Supernova Ia observations to the data analyses, it diminishes (very mildly though) if DES measurements are those included in the combined fit. In this regard, SNIa data is more efficient constraining the neutrino mass, and a $95\%$~CL limit of $0.72$~eV is obtained from the combination of ACT-DR6, BAO and SN data. Notice also from~\autoref{tab:nu} that
the $\sigma_8$ anomaly is absent, as we are only dealing with CMB lensing data and not with CMB temperature observations, which are those driving this tension. This result will remain unchanged in the following sections. 
\autoref{tab:nututto} shows the impact of the addition of Planck lensing measurements to the baseline ACT-DR6 data. Notice first of all that the neutrino mass constraints are notably improved. Also, compared to the results of Planck, the limit $\sum m_\nu<0.60$~eV obtained from the combination of Planck lensing plus BAO and acoustic scale priors quoted in Ref.~\cite{Planck:2018lbu} tightens to $\sum m_\nu<0.49$~eV for ACT-DR6, Planck lensing and BAO and SN priors. 
This constraint loosens when considering DES observations, due to the lower value of $\sigma_8$ preferred by the former dataset and its anti-correlation with $\sum m_\nu$. Concerning the Hubble parameter, the value is always slightly higher than that inferred from CMB temperature anisotropies and therefore the Hubble tension is slightly alleviated. All the constraints shown in ~\autoref{tab:nututto} are illustrated in ~\autoref{fig:nu} (right panel), which shows the posterior probability distributions and the two-dimensional $68\%$ and $95\%$~CL allowed contours.  A very interesting aspect to notice from this figure is the change in the direction of the degeneracy line between $\Omega_{\rm m}$ and $H_0$ when Planck CMB lensing information is added to the baseline ACT-DR6 dataset, as we can see by comparing the difference in the contours in the ($H_0$, $\Omega_{\rm m}$) plane between the left and right panels of ~\autoref{fig:nu}. The reason for that is due to the fact that CMB lensing and BAO constraints on the former plane are almost orthogonal~\cite{ACT:2023kun}. In the case of CMB lensing $\Omega_{\rm m}$ and $H_0$ are anticorrelated, providing ACT-DR6 lensing data the
following constraint on the three-dimensional $\sigma_8$-$H_0$–$\Omega_{\rm m}$ plane~\cite{ACT:2023dou}:
\begin{equation}
\left(\frac{\sigma_8}{0.3}\right)
\left(\frac{\Omega_{\rm m}}{0.3}\right)^{0.23}
\left(\frac{\Omega_{\rm m}h^2}{0.13}\right)^{-0.32}
= 0.994 \pm 0.020~,
\end{equation}
where the error refers to $68\%$~\rm{CL}.
This line-like degeneracy translates into a  narrow region in the $\sigma_8$–$\Omega_{\rm m}$ with 
\begin{equation}
\sigma_8 \Omega_{\rm m}^{0.25}= 0.606 \pm 0.016~,
\end{equation}
with $68\%$~\rm{CL} error. Planck lensing measurements also provide a constraint on a narrow band in the 3-dimensional $\sigma_8$-$H_0$–$\Omega_{\rm m}$ parameter space~\cite{Planck:2018lbu},
\begin{equation}
\left(\frac{\sigma_8}{0.3}\right)
\left(\frac{\Omega_{\rm m}}{0.3}\right)^{0.23}
\left(\frac{\Omega_{\rm m}h^2}{0.13}\right)^{-0.32}
= 0.986 \pm 0.020~,
\end{equation}
with $68\%$~\rm{CL} error. The corresponding band in the $\sigma_8$–$\Omega_{\rm m}$ plane in this case is given by~\cite{Planck:2018lbu}
\begin{equation}
\sigma_8 \Omega_{\rm m}^{0.25}= 0.589 \pm 0.020~,
\end{equation}
with, again, $68\%$~\rm{CL} uncertainty.
In the case of BAO instead,  $\Omega_{\rm m}$ and $H_0$  are positively correlated. Since the BAO constraint dominates that of ACT-DR6, when these two datasets are combined the resulting denegeracy line in the ($H_0$, $\Omega_{\rm m}$) plane follows the trend of the BAO-only contours. However, when considering also Planck CMB lensing measurements the combination of ACT-DR6 with those makes CMB lensing more powerful than BAO observations, following now the degeneracy line in the ($H_0$, $\Omega_{\rm m}$) plane the CMB lensing one, rather than the one from BAO observations.

\subsection{Constraints on Axions}
\label{sec.axion}

\begin{table*}
\begin{center}
\renewcommand{\arraystretch}{1.5}
\resizebox{\textwidth}{!}{
\begin{tabular}{l c c c c c c c c c c c c c c c }
\hline
\textbf{Parameter} & \textbf{ ACT-DR6 } & \textbf{ ACT-DR6 + BAO } & \textbf{ ACT-DR6 + BAO + DES } & \textbf{ ACT-DR6 + BAO + SN } & \textbf{ ACT-DR6 + BAO + DES + SN } \\ 
\hline\hline

$ m_\mathrm{a} \, [eV]  $ & $ < 3.11 $ & $ < 2.19 $ & $ < 1.28 $ & $ < 1.46 $ & $ < 1.27 $ \\ 
$ \Omega_\mathrm{m}  $ & $  0.86^{+0.49}_{-1.1}\, (< 2.86 ) $ & $  0.333^{+0.028}_{-0.032}\, ( 0.333^{+0.061}_{-0.057} ) $ & $  0.293\pm 0.011\, ( 0.293^{+0.023}_{-0.023} ) $ & $  0.314\pm 0.016\, ( 0.314^{+0.033}_{-0.030} ) $ & $  0.294\pm 0.010\, ( 0.294^{+0.020}_{-0.019} ) $ \\ 
$ H_0  $ & $  56^{+20}_{-20} $ (unc) & $  70.2\pm 1.7\, ( 70.2^{+3.6}_{-3.5} ) $ & $  68.5\pm 1.1\, ( 68.5^{+2.1}_{-2.0} ) $ & $  69.4\pm 1.3\, ( 69.4^{+2.6}_{-2.4} ) $ & $  68.5\pm 1.1\, ( 68.5^{+2.1}_{-2.0} ) $ \\ 
$ \sigma_8  $ & $  0.70^{+0.17}_{-0.15}\, ( 0.70^{+0.31}_{-0.33} ) $ & $  0.808\pm 0.017\, ( 0.808^{+0.037}_{-0.037} ) $ & $  0.793\pm 0.015\, ( 0.793^{+0.028}_{-0.031} ) $ & $  0.807\pm 0.018\, ( 0.807^{+0.037}_{-0.038} ) $ & $  0.793\pm 0.015\, ( 0.793^{+0.028}_{-0.030} ) $ \\ 

\hline \hline
\end{tabular} }
\end{center}
\caption{ \small  \textbf{Axion:} Mean values for some of the most relevant cosmological parameters in this study, together with their 68$\%$ (95$\%$) CL errors for the some of the possible data combinations here considered, based on the baseline ACT-DR6  dataset. Upper bounds are quoted at $95\%$~CL significance.}
\label{tab:axion}
\end{table*}

\begin{table*}
\begin{center}
\renewcommand{\arraystretch}{1.5}
\resizebox{\textwidth}{!}{
\begin{tabular}{l c c c c c c c c c c c c c c c }
\hline
\textbf{Parameter} & \textbf{ Full lensing } & \textbf{ Full lensing + BAO } & \textbf{ Full lensing + BAO + DES } & \textbf{ Full lensing + BAO + SN } & \textbf{ Full lensing + BAO + DES + SN } \\ 
\hline\hline

$ m_\mathrm{a} \, [eV]  $ & $ < 1.79 $ & $ < 1.34 $ & $ < 1.61 $ & $ < 1.20 $ & $ < 1.47 $ \\ 
$ \Omega_\mathrm{m}  $ & $  0.382\pm 0.049\, ( 0.382^{+0.10}_{-0.096} ) $ & $  0.3163\pm 0.0088\, ( 0.316^{+0.019}_{-0.019} ) $ & $  0.3102\pm 0.0085\, ( 0.310^{+0.018}_{-0.018} ) $ & $  0.3138\pm 0.0080\, ( 0.314^{+0.016}_{-0.015} ) $ & $  0.3089\pm 0.0078\, ( 0.309^{+0.016}_{-0.014} ) $ \\ 
$ H_0  $ & $  63.4^{+3.7}_{-4.4}\, ( 63^{+8}_{-8} ) $ & $  69.14\pm 0.88\, ( 69.1^{+1.7}_{-1.8} ) $ & $  69.68\pm 0.87\, ( 69.7^{+1.8}_{-1.9} ) $ & $  69.23\pm 0.87\, ( 69.2^{+1.7}_{-1.7} ) $ & $  69.70\pm 0.88\, ( 69.7^{+1.8}_{-1.9} ) $ \\ 
$ \sigma_8  $ & $  0.762\pm 0.031\, ( 0.762^{+0.060}_{-0.061} ) $ & $  0.801\pm 0.016\, ( 0.801^{+0.033}_{-0.035} ) $ & $  0.783\pm 0.015\, ( 0.783^{+0.031}_{-0.032} ) $ & $  0.803\pm 0.015\, ( 0.803^{+0.031}_{-0.032} ) $ & $  0.784\pm 0.014\, ( 0.784^{+0.027}_{-0.029} ) $ \\ 

\hline \hline
\end{tabular} }
\end{center}
\caption{\small \textbf{Axion:} Mean values for some of the most relevant cosmological parameters in this study, together with their 68$\%$ (95$\%$) CL errors for the some of the possible data combinations here considered, based on the baseline ACT-DR6 plus Planck lensing datasets. Upper bounds are quoted at $95\%$~CL significance.}
\label{tab:axiontutto}
\end{table*}

\begin{figure*}[htbp!]
    \begin{tabular}{cc}
   \includegraphics[width=0.5\textwidth]{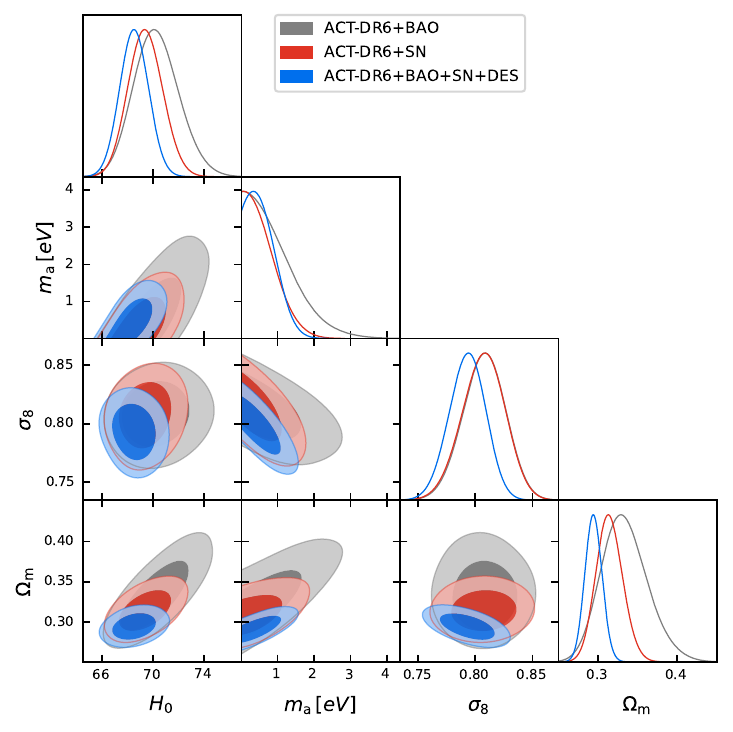}&
   \includegraphics[width=0.5\textwidth]{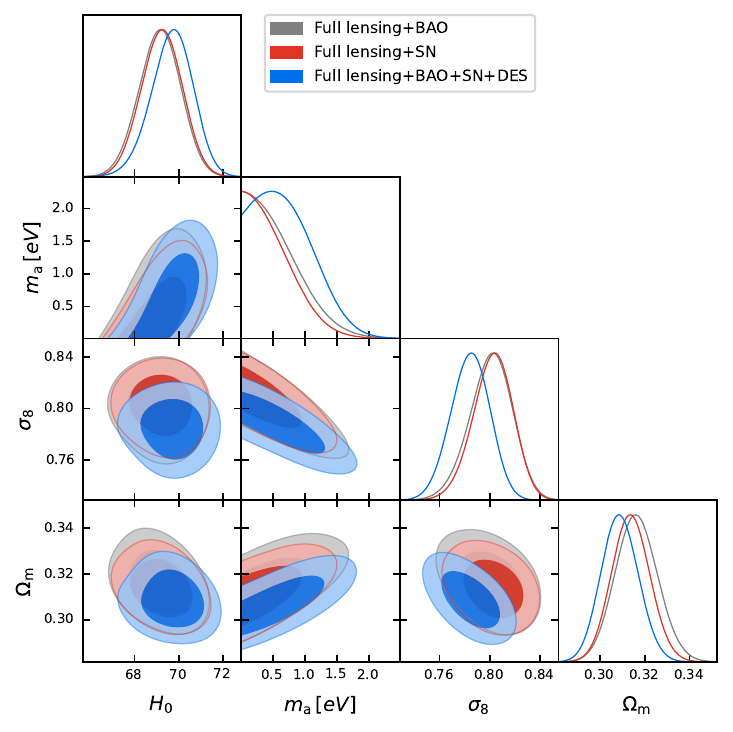}
    \end{tabular}
   \caption{\small \textbf{Axion:} Left (right) panel: One-dimensional posterior probability distributions and two-dimensional $68\%$ and $95\%$~CL allowed contours for $m_a$, $H_0$, $\Omega_{\rm m}$ and $\sigma_8$ from combinations of the baseline ACT-DR6 (ACT-DR6 plus Planck lensing) with other low redshift observables considered along this work.}
    \label{fig:axion}
    
\end{figure*}

The constraints on massive axions, fixing the neutrino mass to $0.06$~eV, are shown in~\autoref{tab:axion} and \autoref{tab:axiontutto} and in~\autoref{fig:axion}. Notice that the results when only ACT-DR6 lensing measurements are considered are very similar to those previously presented for the massive neutrino case, except for a small difference. In the axion case, the addition of DES observations to ACT-DR6 and BAO measurements results in a stronger axion mass bound than that found when SNIa luminosity distance data are considered. The limit is $m_a<1.28$~eV ($m_a<1.46$~eV) at $95\%$~CL for the combination of ACT-DR6 plus BAO plus DES (ACT-DR6 plus BAO plus SN). 
When adding Planck lensing information, the limits on the axion mass are only improved when a few data sets (e.g. lensing plus BAO) are involved, while they are only mildly better for the cases in which a larger sample of observations are analysed. For instance, from the combination of ACT-DR6 plus BAO plus SN the $95\%$~CL upper bound on the axion mass changes from $1.46$~eV to $1.2$~eV when adding Planck lensing data. As can be noticed from ~\autoref{fig:axion} also for this case, as in the massive neutrino one, there is a change in the direction of the degeneracy line between $\Omega_{\rm m}$ and $H_0$ when Planck CMB lensing information is added to the baseline ACT-DR6 dataset. The values of the Hubble constant are also always slightly higher than those obtained in the canonical $\Lambda$CDM scenario with Planck temperature anisotropies: for the combination of ACT-DR6 plus BAO, we obtain $H_0=70.2\pm 1.7$, which is much closer to the value measured by local probes, $H_0=73.04 \pm 1.04$ km/s/Mpc~\cite{Riess:2021jrx}, lowering considerably the statistical significance of the so-called Hubble constant tension. A similar argument applies to almost all the remaining cases when considering ACT-DR6 plus Planck lensing data combined.

\begin{table*}
\begin{center}
\renewcommand{\arraystretch}{1.5}
\resizebox{\textwidth}{!}{
\begin{tabular}{l c c c c c c c c c c c c c c c }
\hline
\textbf{Parameter} & \textbf{ ACT-DR6 } & \textbf{ ACT-DR6 + BAO } & \textbf{ ACT-DR6 + BAO + DES } & \textbf{ ACT-DR6 + BAO + SN } & \textbf{ ACT-DR6 + BAO + DES + SN } \\ 
\hline\hline

$ m_\mathrm{a} \, [eV]  $ & $ < 4.06 $ & $ < 1.86 $ & $ < 1.07 $ & $ < 1.19 $ & $ < 1.08 $ \\ 
$ \sum m_\nu \, [eV]  $ & $ < 3.67 $ & $ < 1.14 $ & $ < 0.732 $ & $ < 0.654 $ & $ < 0.684 $ \\ 
$ \Omega_\mathrm{m}  $ & $  1.7^{+1.1}_{-2.1}\, (\rm{unc}) $ & $  0.359\pm 0.033\, ( 0.359^{+0.073}_{-0.069} ) $ & $  0.306\pm 0.013\, ( 0.306^{+0.027}_{-0.025} ) $ & $  0.320\pm 0.016\, ( 0.320^{+0.035}_{-0.034} ) $ & $  0.305\pm 0.012\, ( 0.305^{+0.025}_{-0.024} ) $ \\ 
$ H_0  $ & $ < 55.5 \, (\rm{unc})$ & $70.6\pm 1.8\, ( 70.6^{+3.8}_{-3.7} ) $ & $  68.3\pm 1.0\, ( 68.3^{+2.0}_{-1.9} ) $ & $  69.1\pm 1.2\, ( 69.1^{+2.7}_{-2.6} ) $ & $  68.3\pm 1.0\, ( 68.3^{+2.1}_{-2.0} ) $\\ 
$ \sigma_8  $ & $  0.57\pm 0.16\, ( 0.57^{+0.35}_{-0.34} ) $ & $  0.792\pm 0.019\, ( 0.792^{+0.037}_{-0.038} ) $ & $  0.774\pm 0.017\, ( 0.774^{+0.033}_{-0.034} ) $ & $  0.792\pm 0.020\, ( 0.792^{+0.037}_{-0.038} ) $ & $  0.775\pm 0.017\, ( 0.775^{+0.032}_{-0.032} ) $ \\ 

\hline \hline
\end{tabular} }
\end{center}
\caption{ \small \textbf{Axion $\&$ Neutrinos:} Mean values for some of the most relevant cosmological parameters in this study, together with their 68$\%$ (95$\%$) CL errors for the some of the possible data combinations here considered, based on the baseline ACT-DR6 lensing dataset. Upper bounds are quoted at $95\%$~CL significance. }
\label{tab:joint}
\end{table*}

\begin{table*}
\begin{center}
\renewcommand{\arraystretch}{1.5}
\resizebox{\textwidth}{!}{
\begin{tabular}{l c c c c c c c c c c c c c c c }
\hline
\textbf{Parameter} & \textbf{ Full lensing } & \textbf{ Full lensing + BAO } & \textbf{ Full lensing + BAO + DES } & \textbf{ Full lensing + BAO + SN } & \textbf{ Full lensing + BAO + DES + SN } \\ 
\hline\hline

$ m_\mathrm{a} \, [eV]  $ & $ < 1.79 $ & $ < 1.23 $ & $ < 1.38 $ & $ < 1.11 $ & $ < 1.32 $ \\ 
$ \sum m_\nu \, [eV]  $ & $ < 1.32 $ & $ < 0.492 $ & $ < 0.583 $ & $ < 0.432 $ & $ < 0.533 $ \\ 
$ \Omega_\mathrm{m}  $ & $  0.56^{+0.12}_{-0.14}\, ( 0.56^{+0.28}_{-0.26} ) $ & $  0.324\pm 0.010\, ( 0.324^{+0.022}_{-0.022} ) $ & $  0.319\pm 0.010\, ( 0.319^{+0.022}_{-0.021} ) $ & $  0.3191\pm 0.0090\, ( 0.319^{+0.018}_{-0.017} ) $ & $  0.3154\pm 0.0091\, ( 0.315^{+0.018}_{-0.017} ) $ \\ 
$ H_0  $ & $  55.8\pm 5.0\, ( 56^{+10}_{-10} ) $ & $  69.01\pm 0.86\, ( 69.0^{+1.7}_{-1.7} ) $ & $  69.51\pm 0.85\, ( 69.5^{+1.8}_{-1.8} ) $ & $  69.17\pm 0.85\, ( 69.2^{+1.7}_{-1.7} ) $ & $  69.61\pm 0.83\, ( 69.6^{+1.6}_{-1.7} ) $ \\ 
$ \sigma_8  $ & $  0.682\pm 0.047\, ( 0.682^{+0.091}_{-0.087} ) $ & $  0.789\pm 0.018\, ( 0.789^{+0.033}_{-0.036} ) $ & $  0.770\pm 0.016\, ( 0.770^{+0.033}_{-0.032} ) $ & $  0.792\pm 0.017\, ( 0.792^{+0.035}_{-0.036} ) $ & $  0.774\pm 0.015\, ( 0.774^{+0.029}_{-0.031} ) $ \\ 

\hline \hline
\end{tabular} }
\end{center}
\caption{ \small \textbf{Axion $\&$ Neutrinos:} Mean values for some of the most relevant cosmological parameters in this study, together with their 68$\%$ (95$\%$) CL errors for the some of the possible data combinations here considered, based on the baseline ACT-DR6 plus Planck lensing datasets. Upper bounds are quoted at $95\%$~CL significance. }
\label{tab:jointtutto}
\end{table*}

\begin{figure*}[ht!]
\begin{tabular}{c}
\includegraphics[width=0.5\textwidth]{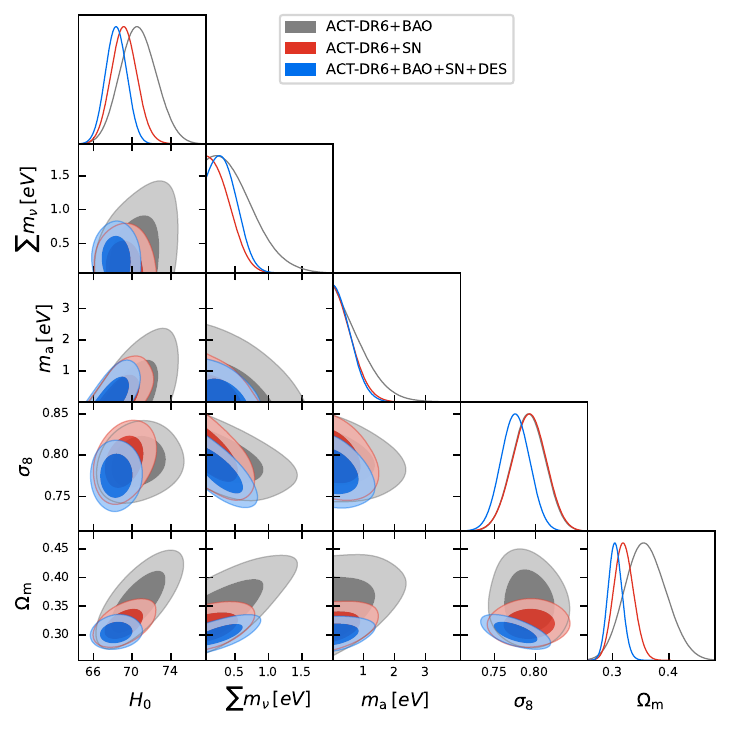} 
\includegraphics[width=0.5\textwidth]{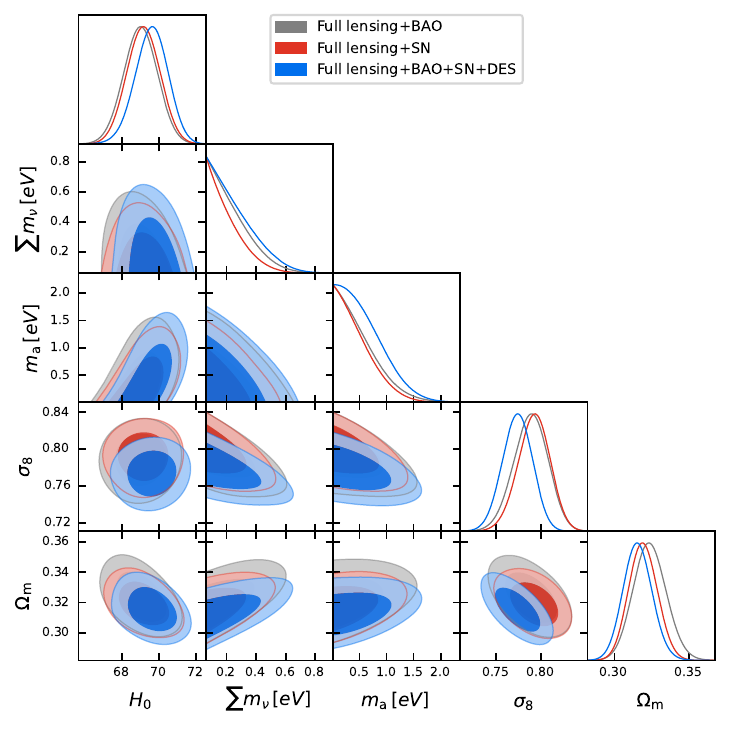}
\end{tabular}
\caption{\small \textbf{Axion $\&$ Neutrinos:} Left (right) panel: One-dimensional posterior probability distributions and two-dimensional $68\%$ and $95\%$~CL allowed contours for $\sum m_\nu$, $m_a$, $H_0$, $\Omega_{\rm m}$ and $\sigma_8$ from combinations of the baseline ACT-DR6 (ACT-DR6 plus Planck lensing) with other low redshift observables considered along this work.}
\label{fig:joint}
\end{figure*}


\subsection{Joint Constraints on Axions \& Neutrinos}
\label{sec.joint}
The constraints on a mixed dark matter scenario are summarized in ~\autoref{tab:joint}, \autoref{tab:jointtutto} and ~\autoref{fig:joint}. 
We notice that the limits on both the neutrino and the axion masses shown in~\autoref{tab:joint}, even if close to those obtained in the neutrino-only and axion-only hot dark matter scenarios (see~\autoref{tab:nu} and \autoref{tab:axion}), are more constraining, due to the fact that the hot dark matter energy density is now shared among two different species and therefore the amount of each of them is reduced with respect to the case in which only one of them is present. The very same argument applies when considering also Planck lensing measurements in the data analyses. The tightest $95\%$~CL limits we find here are $\sum m_\nu<0.43$~eV and $m_a<1.11$~eV for the combination of CMB lensing data, BAO and SN observations. When compared to the tightest cosmological limits quoted in the literature, the former constraints may seem not highly competitive, but they are impressively robust and independent, as they do not rely on CMB temperature and polarization anisotropies: they are based on low redshift phenomena as lensing and large scale structure data. As in the previous cases, within this mixed hot dark matter scenario the Hubble constant tension is also relieved, i.e. the significance is much smaller than in the standard $\Lambda$CDM case. In some cases, it does not neither reach the $2.5\sigma$ significance level.

\clearpage
\section{Conclusions}
\label{sec.conclusion}
Low-redshift phenomena play a highly relevant role in cosmology nowadays. Lensing of the Cosmic Microwave Background photons and of large scale structure by intervening galaxies, as well as large scale structure standard rulers (BAO) have been shown to improve the bounds on a variety of cosmological parameters derived from Planck CMB temperature and polarization anisotropies. Very well-known examples are the neutrino and the axion masses and abundances. The constraining power of cosmological observations is much more efficient when these low redshift observations are included in the data analyses. However, a pending question is how robust would be the limits that these low-redhsift probes impose by themselves, without relying in the CMB temperature input. Here we have explored such a situation, finding very competitive limits, much superior than those found in current laboratory searches for thermal relic properties. The tightest bounds we find for the neutrino and axion masses are $\sum m_\nu < 0.43$~eV and $m_a<1.1$~eV at 95\% CL respectively, for the Full lensing + BAO + SN dataset combination.

These limits reassess both the robustness and the constraining power of cosmological thermal relic searches. For instance, in the neutrino case, the limits inferred by current beta-decay experiments such as KATRIN impose that $\sum m_\nu \lesssim 2.0$~eV~\cite{KATRIN:2021uub}, while the bounds on the effective Majorana neutrino mass from neutrinoless double beta decay experiments set the bound $\sum m_\nu \lesssim 0.1-2.0$~eV~\cite{KamLAND-Zen:2022tow}, both at $90\%$~CL. 

Intriguingly, also when discarding CMB temperature and polarization anistropies in the data analyses, long-standing tensions such as the Hubble constant and the clustering parameter $\sigma_8$ ones are either much less significant or copletely absent. Future lensing measurements 
from CMB and/or galaxy surveys may have the key to resolve these pending issues. 

\begin{acknowledgments}

This work has been supported by the Spanish MCIN/AEI/10.13039/501100011033 grants PID2020-113644GB-I00 (RH and OM) and by the European ITN project HIDDeN (H2020-MSCA-ITN-2019/860881-HIDDeN) and SE project ASYMMETRY (HORIZON-MSCA-2021-SE-01/101086085-ASYMMETRY) and well as by the Generalitat Valenciana grants PROMETEO/2019/083 and CIPROM/2022/69. EDV is supported by a Royal Society Dorothy Hodgkin Research Fellowship. This article is based upon work from COST Action CA21136 Addressing observational tensions in cosmology with systematics and fundamental physics (CosmoVerse) supported by COST (European Cooperation in Science and Technology).
We acknowledge IT Services at The University of Sheffield for the provision of services for High Performance Computing.
\end{acknowledgments}

\bibliography{bib}
\end{document}